\documentclass[sigconf,nonacm]{acmart}
\setcopyright{none}
\renewcommand\footnotetextcopyrightpermission[1]{}
\usepackage{multirow}
\usepackage{tabularx}
\usepackage{array}
\usepackage{xurl}
\newcommand{\bmhead}[1]{\paragraph{#1.}}

\newcommand{\degr}{\ensuremath{^\circ}}

\begin{document}
\title{Exact Embedding Verification on an Edge NPU: Measurements and Deployment Constraints}

\author{Georgios Feretzakis}
\affiliation{\institution{securagen.ai}\city{Athens}\country{Greece}}
\email{research@securagen.ai}
\author{Alexandros Papaspyridis}
\affiliation{\institution{securagen.ai}\city{Athens}\country{Greece}}

\begin{abstract}
Offline retrieval-augmented generation (RAG) on edge devices receives knowledge as signed packs of text and precomputed embedding vectors. Prior work showed that a signature does not bind the vectors to the text, that any verifier which accepts recomputed vectors within a cosine tolerance concedes a predictable share of queries, and that on a Raspberry Pi~5 the recomputation needed for exact checks costs a concurrently running language model 23--31\% of its decoding speed on the CPU. Here we ask whether a low-cost neural processing unit (NPU) makes exact verification practical. On a Raspberry Pi~5 with a Hailo-8L accelerator, verifying 8 text--vector pairs per second costs the language model 1.3\% of its decoding speed instead of 23--31\%, and re-embedding a 1,745-window corpus takes 3.4~s instead of 72.9~s in FP32 on the CPU. The NPU's INT8 embeddings deviate from floating point (median cosine 0.917), so a tolerance set at the empirical 0.1\% quantile of honest NPU--FP32 cosine similarities (0.784, about 38\degr) would let an attacker who controls the vectors displace the correct document for essentially all evaluated queries; with that tolerance applied as an angular budget in a separate BGE-M3 experiment, an end-to-end attack lowers correct answers from 0.75 to 0.35. The same chip, however, returns bit-identical embeddings across repetitions, batch sizes and processes, which makes exact comparison against a reference generated on the same tested device and configuration possible. We also describe how to deploy exact verification: the device recomputes every chunk, retrieves only with its own vectors and, where the reference comes from a computation shown to reproduce its output, compares shipped reference vectors bit for bit; a mismatch is treated as an integrity event to be investigated rather than resolved by a tolerance; and a signed manifest binds every element that determines the bits.
\end{abstract}



\maketitle

\section{Introduction}\label{sec:intro}
Retrieval-augmented generation (RAG) running fully offline on small boards~\cite{prior}, for example with llama.cpp and Qwen2.5~\cite{r9,r104}, receives its knowledge through update packages. Signed, versioned knowledge packs~\cite{prior} authenticate who published an update, but the study \emph{Signed but Steerable}~\cite{sbs} showed that this is not enough when packs ship precomputed embedding vectors with their text: a signature proves who published the vectors, not that they encode the text. That study derived the \emph{margin law}, which states that a verifier accepting a recomputed vector within cosine tolerance $\tau$ leaves exactly the queries whose angular retrieval margin is below $\arccos\tau$ open to an attacker who controls the vectors, and showed that the gap closes only when the publisher's vectors play no role in retrieval: by recomputing and discarding them on the device, by exact comparison with a bit-reproducible computation, or, for compressed stores, by code-level verification with canonicalisation~\cite{sbs}.

That analysis left a practical question open. On the CPU of a Raspberry Pi~5, the recomputation that these designs require competes with the language model: verifying 8 text--vector pairs per second during generation cost the model 23--31\% of its decoding speed~\cite{sbs}. And exact comparison needs a computation that reproduces itself bit for bit, which none of the floating-point or dynamic-INT8 pipelines measured in~\cite{sbs} did. Low-cost neural processing units (NPUs) now sit next to such boards. They could take the recomputation off the CPU, but they compute in quantised integer arithmetic, whose outputs deviate from floating point. This paper asks whether an edge NPU makes exact embedding verification practical, and what it costs.

We evaluate a Hailo-8L on a Raspberry Pi~5 and report three results and a deployment design:
\begin{enumerate}
\item \textbf{Cost} (Sect.~\ref{sec:cost}): verifying 8 text--vector pairs per second costs the concurrently running language model 1.3\% of its decoding speed on the NPU, against 23--31\% on the CPU, with the 99th-percentile inter-token latency almost unchanged; re-embedding a 1,745-window corpus takes 3.4~s instead of 72.9~s in FP32 on the CPU.
\item \textbf{Determinism} (Sect.~\ref{sec:det}): the chip returns bit-identical embeddings across repetitions, batch sizes and processes.
\item \textbf{The tolerance trap} (Sect.~\ref{sec:fidelity}): the NPU's INT8 embeddings are good enough for retrieval but deviate from floating point; a tolerance-based check wide enough for them would, by the margin law, leave essentially all queries exposed.
\item \textbf{Deployment} (Sect.~\ref{sec:deploy}): a design for exact verification on a deterministic device, which combines recomputation with exact comparison, states who can produce the reference, what a signed manifest must bind, and how a device handles a mismatch. This part is a design and contains no new measurements.
\end{enumerate}
Together they say that the quantised execution which defeats tolerance-based checks against floating-point vectors, for an attacker who controls the vectors, is also what makes exact verification practical on the deployed device. The study~\cite{sbs} measured only general-purpose CPUs and GPUs and left dedicated accelerators to future work; this paper takes up that question.

\emph{Related work.} Floating-point non-associativity and kernels that are not invariant to batch composition make inference results depend on how a request is executed~\cite{he2025}, and dynamic quantisation can couple co-batched requests~\cite{quantamination}. Verifiable-inference systems handle this either with tolerances, using locality-sensitive commitments to intermediate activations~\cite{toploc}, or by removing nondeterminism, with pinned kernels and drivers on a fixed GPU architecture~\cite{eigenai} or with integer-only arithmetic~\cite{barry2026,intllm} building on integer-only quantisation~\cite{jacob2018,ibert}. Prior work transferred the tolerance-versus-exactness question to retrieval vectors~\cite{sbs}. Here we examine an off-the-shelf edge accelerator whose bit-identical output makes exact comparison possible on the device, and quantify what it costs next to a running language model.

\section{Setting and Method}\label{sec:method}
\emph{Threat model.} We follow~\cite{sbs}: a publisher, possibly compromised, ships a signed pack of text chunks and vectors; every chunk's text passes content review, and the adversary chooses the vectors. The device can recompute embeddings of the text with the agreed model and configuration, which the adversary does not control.

\emph{Hardware and software.} Raspberry Pi~5 (16~GB) with the Raspberry Pi AI HAT+ (Hailo-8L, 13~TOPS, PCIe Gen3 x1)~\cite{aihat}; HailoRT~4.23.0. The language model is Qwen2.5-3B-Instruct (Q4\_K\_M) under llama.cpp on the Pi's CPU~\cite{r9,r104}, with the build and settings of~\cite{sbs}.

\emph{Model and data.} The embedding model is all-MiniLM-L6-v2~\cite{sbert,minilm}; the CPU and FP32 computations use the checkpoint \nolinkurl{sentence-transformers/all-MiniLM-L6-v2} at revision \texttt{1110a243}~\cite{hfminilm}. Retrieval and abstention use the 300-question set of~\cite{prior,sbs} (100 answerable, 100 in-topic unanswerable and 100 out-of-corpus questions): recall@3 and the share of answerable questions retained are computed over the 100 answerable questions, and the abstention AUROC compares the 100 answerable with the 100 in-topic unanswerable questions. We use the compiled graph from the Hailo Model Zoo v2.19.0 (\nolinkurl{all_minilm_l6_v2}, HAILO8L, 128-token windows)~\cite{hailozoo}. The corpus consists of 1,745 windows of up to 128 tokens built from SQuAD~2.0~\cite{r105}, and the question set is that of~\cite{prior,sbs}; the NPU and FP32 embeddings compared in Sect.~\ref{sec:fidelity} are computed for the same windows. The margin-law analysis of Sect.~\ref{sec:fidelity} uses the five corpora (SQuAD~2.0 and the BEIR~\cite{beir} corpora SciFact, NFCorpus, ArguAna and FiQA) and the two embedding models (BGE-M3~\cite{bgem3} and all-MiniLM-L6-v2) of~\cite{sbs}. Floating-point (FP32) references are computed with ONNX Runtime on the Pi's CPU; CPU baselines use ONNX Runtime with dynamic INT8 quantisation. The CPU figures for re-embedding the corpus and for the cost of CPU verification next to the language model (Table~\ref{tab:contention}) are those reported in~\cite{sbs} for a Raspberry Pi~5 with the same corpus and language model.

\emph{Measurement conditions.} The cost of verification next to the language model is the mean over three full runs, with the encoder either saturating the CPU or the NPU (re-indexing) or running periodically at 8 text--vector pairs per second; the reference without an encoder decodes 5.14 tokens/s with a 99th-percentile (p99) inter-token latency of 210~ms. Latencies are medians. The end-to-end experiment follows~\cite{sbs}: for the same 100 answerable questions, one instruction-bearing chunk is added to the pack, and its shipped vector is rotated towards the query, within the angular budget $\arccos\tau$, in the BGE-M3 space of the knowledge pack of~\cite{prior,sbs}, so that it enters the top-3 context; the tolerance measured for all-MiniLM-L6-v2 on the NPU is applied as this angular budget, as~\cite{sbs} did for the CPU-INT8 tolerance.

\emph{Disclosure.} Hailo granted written permission to publish the Hailo-8L results in this paper. They are not official Hailo benchmarks and do not imply endorsement by Hailo; Hailo had no role in the design of the study.

\section{Results}\label{sec:results}

\subsection{Throughput and Cost Next to the Language Model}\label{sec:cost}
At batch~1 the NPU's median latency is 14.3~ms per window, the same as the fastest CPU configuration (dynamic INT8, four threads, 14.3~ms), which uses four CPU cores against 0.12 for the NPU. From batch~2 the NPU's throughput exceeds every CPU configuration we measured (131 embeddings/s against at most 74). It reaches 517 embeddings/s at batch~32 using 0.47 CPU cores, and 575 embeddings/s at batch~63, the largest batch tested. Re-indexing the 1,745-window corpus takes 3.4~s on the NPU (0.48 CPU cores) against 72.9~s in FP32 on four CPU cores.

Table~\ref{tab:contention} reports the cost to the language model. Verification during generation costs the LLM 1.3\% of its decoding speed on the NPU and 23--31\% on the CPU, where it also raises the 99th-percentile inter-token latency by 108--167\%.

\begin{table*}[t]
\centering\small
\caption{Local LLM decoding on the Raspberry Pi~5 with an embedding encoder running concurrently. Mean [range] over three full runs; CPU rows give the range over 1, 2 and 4 threads of the three-run means; encoder rates are three-run means. Reference without encoder: 5.14 tokens/s, p99 inter-token latency 210~ms}\label{tab:contention}
\begin{tabularx}{\textwidth}{lXX}
\toprule
Concurrent encoder & Saturating (re-indexing) & Periodic verification, 8 pairs/s\\
\midrule
CPU INT8, 1--4 threads & decode $-64.5\%$ to $-73.9\%$ (16--25 items/s) & decode $-22.9\%$ to $-31.4\%$; p99 inter-token latency $+108\%$ to $+167\%$\\
Hailo-8L & decode $-35.6\%$ [35.4, 35.9] (331 items/s) & decode $-1.3\%$ [1.1, 1.5]; p99 $+4\%$ [3, 6]\\
\bottomrule
\end{tabularx}
\end{table*}

\subsection{Determinism}\label{sec:det}
With the same 256 corpus windows, the Hailo-8L returns bit-identical embeddings across five repetitions, batch sizes 1, 8 and 32, and a fresh process. This contrasts with the floating-point pipelines measured on CPUs and GPUs in~\cite{sbs}, which varied with batch size or architecture, and with dynamic INT8 on the CPU, whose outputs depend on the other inputs in the batch~\cite{sbs,quantamination}. Bit-identity across \emph{different} chips of the same model was not tested.

\subsection{Fidelity and the Tolerance Trap}\label{sec:fidelity}
Over the 1,745 windows, the median cosine between NPU and FP32 embeddings is 0.917 (minimum 0.769). For retrieval this deviation makes little difference on our data: retrieval recall@3 is 0.85 on the NPU versus 0.87 for the same model in FP32; for the relevance-based abstention scorer of~\cite{sbs} (question--sentence similarity), the AUROC is 0.692 [0.614, 0.766] on the NPU versus 0.690 [0.614, 0.763] in FP32 (95\% intervals), and a threshold tuned in FP32 retains 0.93 of answerable questions on the NPU versus 0.91 in FP32.

For verification it matters a great deal. A verifier that compares NPU embeddings with FP32 references must tolerate the NPU's deviation. The tolerance such a verifier needs for honest content, the 0.1\% quantile of honest NPU--FP32 cosines, is 0.784 (about 38\degr). Under the margin law~\cite{sbs}, a tolerance this wide lets an attacker who controls the vectors displace the correct document for essentially all queries (100\% point estimate in each of five corpora with both embedding models; lower 95\% bounds 0.97--0.999). In an end-to-end experiment on the device within that tolerance, the language model's correct answers fall from 0.75 to 0.35 and refusals rise from 0.13 to 0.47.

These results concern the design of tolerance-based verification, not the chip. The same mechanism can arise with other quantised accelerators or CPU pipelines; its magnitude depends on the required tolerance and on the retrieval margins. None of our results indicates a defect of the Hailo-8L: deviation from floating point is inherent to quantised inference on any accelerator, and the exposure we measure is a property of the verification design, not of the chip.

\section{Deploying Exact Verification}\label{sec:deploy}
This section turns the measurements into a design. It describes how a deterministic device can use exact verification; it reports no new measurements.

\emph{Verify, then discard.} In~\cite{sbs}, recompute-and-discard and exact comparison appear as alternatives. On a device that recomputes cheaply and deterministically they combine. Before a new chunk becomes searchable, the device embeds it, compares its result bit for bit with the reference vector shipped in the pack and decides whether to accept the pack; it then retrieves only with its own vectors. Retrieval is protected by the recomputation alone, as in recompute-and-discard; the comparison adds detection. Without it, a pack with tampered vectors would be silently repaired and nobody would learn that the publisher had shipped them; with it, the inconsistency becomes an event that can be logged, attributed through the pack's signatures and handled by update policy, the same argument that~\cite{sbs} makes for checking stored codes before canonicalising them. Packs still ship vectors because not every device in a fleet can recompute cheaply; on a device that can, the shipped vectors serve only as the publisher's declared reference, and comparing against them turns an inconsistent pack into a detectable event. Exact verification establishes only that the vectors are consistent with the text under the agreed computation; whether the text itself is correct, authorised or safe is a matter for content review, as in the threat model of~\cite{sbs}. Because the comparison is exact, there is no tolerance to calibrate and, by the margin law, no conceded query.

\emph{Who produces the reference.} Exact comparison is meaningful only if the reference was produced by the same computation as the device. We see three options. (i)~The publisher produces it on a device of the same model, with the same compiled graph and runtime; the device then detects tampered vectors at ingestion. This option, and option~(ii), require that different units of the same model agree bit for bit, which we did not test. (ii)~An independent auditor with such a device produces or re-checks the reference and co-signs its digests, in the spirit of the threshold signatures used for knowledge packs~\cite{prior}; this separates the party that writes content from the party that attests its vectors. (iii)~The device produces the reference itself at first installation and records the digests of its vectors in its local state; later checks of its own store, for instance after a restore from backup, after copying (not recomputing) the store to another unit, or after a storage fault, compare against these digests. The digests need a protected root of trust and rollback protection, for example the signed, monotonic version state already kept for knowledge packs~\cite{prior}; an append-only file alone does not provide this. A unit that receives a copied store can check its bytes against these digests, but this does not re-establish that the vectors encode the text: it relies on the unit that computed them. By recomputing, the receiving unit obtains vectors of its own that it can trust; it can check the copied vectors by exact equality only if its computation is shown to produce the same bytes, which for different chips we did not test. Option~(iii) protects the device's store against later modification but says nothing about the publisher's shipped vectors, which it never uses. Whether a publisher without the device can produce a matching reference by other means is open (Sect.~\ref{sec:discussion}).

\emph{What the manifest binds.} Extending the guidelines of~\cite{sbs} on binding the configuration and pinning everything that affects numerics, a pack that ships reference vectors for exact comparison should bind, in its signed manifest: the model revision and tokeniser; the window length, truncation, padding and pooling rules and the normalisation; the digest of the compiled graph the device executes and the runtime and firmware versions; the accelerator model; and the byte representation on which equality is defined (raw output codes or dequantised values, data type and byte order), together with the hash function applied to it. Equality is then a comparison of byte strings, not of numbers after further processing. The device checks these digests against a configuration policy it obtained independently of the publisher, for example at provisioning: the publisher's signature shows who declared a configuration, not that the configuration is the agreed one, just as a verifier should not accept a codebook from the party whose codes it verifies~\cite{sbs}.

\emph{Conditions under which equality may be relied on.} Bit-identity holds only for the conditions in which it was shown. Our measurements cover repetitions, batch sizes 1, 8 and 32 and a fresh process on one chip (Sect.~\ref{sec:det}). A deployment should test the conditions it actually uses, such as other batch sizes or another unit of the same model, before relying on equality, and should regenerate the reference after any change to an element bound in the manifest, including a runtime or firmware update.

\emph{Handling a mismatch.} A mismatch should never be resolved by falling back to a tolerance: Section~\ref{sec:fidelity} shows that a tolerance wide enough for quantised output concedes essentially every query. Two cases should be distinguished. If the manifest's configuration digests differ from the device's own, the reference was produced by a different computation; the device does not compare, retrieves with its own vectors and reports the configuration difference; a deployment with stricter requirements may instead reject the pack. If the configuration matches and the bits differ, the shipped vectors are inconsistent with the text under the declared computation; this is an integrity event to be investigated. It may indicate tampering, but also storage corruption, an undeclared difference in the execution environment or a condition under which determinism was not tested. The device holds the affected chunks back from the index, records their identifiers and the manifest digest, and applies its update policy, for example quarantining the pack and reporting the discrepancy to the publisher.

\section{Discussion}\label{sec:discussion}
\emph{What changes with an NPU.} In~\cite{sbs} the safe designs had a price on the CPU of a small board. With the NPU, recompute-and-discard becomes affordable during generation: the device can embed every incoming chunk itself and retrieve with its own vectors at a cost of 1.3\% of decoding speed at the tested rate of eight pairs per second; at full re-indexing speed the cost is 35.6\% (Table~\ref{tab:contention}). Exact verification also becomes possible, because the tested chip returned bit-identical outputs under the tested conditions: a reference produced on the same tested device and configuration can be compared bit for bit, with no tolerance and therefore, by the margin law, no conceded queries.

\emph{The tolerance trap.} The NPU's own deviation from floating point is exactly what makes a mixed design dangerous. A verifier that recomputes on the NPU but compares against FP32 vectors shipped by the publisher must accept deviations of about 38\degr, the angle corresponding to the empirical 0.1\% quantile (0.784) of honest NPU--FP32 cosine similarities in our measurements, and then, by the margin law, concedes to an attacker who controls the vectors essentially every evaluated query whose correct document ranks first, in the corpora and models of Sect.~\ref{sec:fidelity}. An empirical quantile is not a guarantee for other data, devices or conditions. The NPU should therefore be used either to recompute and discard, or to compare exactly against NPU-generated references; it should not be used inside a tolerance check against floating-point vectors. The study~\cite{sbs} observed the same effect for dynamic INT8 inference on the CPU; the trap therefore arises from combining a quantised recomputation with floating-point references, wherever the quantised computation runs.

\emph{Who can produce the reference?} Exact comparison requires that whoever produces the reference vectors runs the same computation as the device. A publisher can do so only on hardware whose output is bit-identical to the device's; we showed this for one chip against itself. A publisher without such hardware needs another computation that reproduces the device's output bit for bit; whether such a computation exists, and whether different chips of the same model agree bit for bit, are open questions. Until they are answered, our measurements support only option~(iii) of Sect.~\ref{sec:deploy}, in which the reference is produced on the device that later checks it. In that option the device is protected, because it retrieves only with its own vectors, but it does not verify the vectors shipped by the publisher; detecting a publisher's tampering by exact comparison requires option~(i) or~(ii), and therefore bit-identity across units.

\emph{Limitations.} We measured one board, one Hailo-8L, one embedding model with 128-token windows, one language model and one runtime version. Determinism across chips of the same model, and whether a publisher without the device can reproduce its output, were not studied. The NPU runs the compiled graph of the Hailo Model Zoo, and the FP32 reference runs the original checkpoint exported to ONNX; we did not separate the part of their deviation that is due to quantisation from the part due to other differences between the two pipelines. The tolerance analysis applies the margin law of~\cite{sbs} to the NPU's measured deviation, and the end-to-end experiment applies the tolerance measured for all-MiniLM-L6-v2 as an angular budget in the BGE-M3 retrieval space of~\cite{sbs}; it was not run in the all-MiniLM-L6-v2 retrieval space. The deployment design of Sect.~\ref{sec:deploy} has not been evaluated as a system; the cost measurements of Sect.~\ref{sec:cost} cover the recomputation, not the remaining steps of that design.

\section{Conclusion}\label{sec:conclusion}
The NPU works well as a trust co-processor next to a language model running on the CPU: verifying 8 pairs per second costs the model 1.3\% of its decoding speed against 23--31\% on the CPU, with the 99th-percentile inter-token latency almost unchanged; its determinism allows exact comparison against a reference generated on the same tested device and configuration. Its deviation from floating point, on the other hand, means that a tolerance wide enough for its honest output concedes essentially every evaluated query to an attacker who controls the vectors. Edge accelerators therefore change which integrity checks are practical, provided they are used for recomputation and exact comparison, not for tolerance checks. Used this way, a device can verify, then discard: it retrieves only with its own vectors, compares shipped references bit for bit to detect inconsistent packs when the reference comes from a computation shown to reproduce its output, and treats a mismatch as an integrity event to be investigated rather than as noise.

\section*{Statements and Declarations}
\bmhead{Permission and disclosure} Hailo Technologies Ltd.\ granted written permission to publish the Hailo-8L results reported in this paper. The results are not official Hailo benchmarks and do not imply endorsement by Hailo; Hailo had no role in the design of the study, the analysis or the conclusions.
\bmhead{Funding} This work was funded by securagen.ai.
\bmhead{Competing interests} Georgios Feretzakis is the founder of securagen.ai P.C.; the authors are affiliated with securagen.ai P.C., which develops commercial software in the areas discussed in this paper. The authors have no financial relationship with Hailo Technologies Ltd.
\bmhead{Data availability} The measurements of~\cite{sbs} used here (re-embedding time, cost of CPU verification next to the language model, FP32 retrieval baseline), the margin-law analysis and the code are available with~\cite{sbs} (doi:10.5281/zenodo.23948839). The CPU micro-benchmark data behind the CPU figures of Sect.~\ref{sec:cost} are available at \url{https://github.com/securagenai/edge-embedding-cpu-benchmarks}; the FP32 abstention baseline of Sect.~\ref{sec:fidelity} is available from the authors on request. The Hailo-8L measurements are reported in this paper; the raw measurement files are not released, because they include results outside the scope of the permission to publish.
\bmhead{Trademarks} Hailo and Hailo-8L are trademarks of Hailo Technologies Ltd.; Raspberry Pi is a trademark of Raspberry Pi Ltd.; other product names are trademarks of their respective owners.

\bmhead{Author contribution} Georgios Feretzakis designed the methodology, conducted the experiments, analysed the results and wrote the manuscript. Alexandros Papaspyridis critically revised the work. Both authors read and approved the final manuscript.

\section*{GenAI Usage Disclosure}
The authors used standard writing and production tools for language editing and \LaTeX\ preparation. Generative AI assistants were used to help identify candidate literature, to cross-check bibliographic metadata, and to help write the experiment and analysis scripts; every reference was verified by the authors against the publisher, DOI, or preprint-server record, and all code and results were reviewed by the authors. The AI tools did not generate the framework, the findings, or the conclusions. The authors take full responsibility for the manuscript.

\bibliographystyle{ACM-Reference-Format}
\bibliography{references}
\end{document}